\documentclass[11pt,a4paper]{elsart}%letter
\usepackage{ifthen,graphics}
\usepackage{color}
\usepackage{cite}
\usepackage{amssymb}
\usepackage{epsfig}
\usepackage[colorlinks]{hyperref}

\definecolor{dgreen}{rgb}{.0,.7,.0}

\newcommand{\Eq}[1]{Eq.~(\ref{#1})}

\newcommand{\Fig}[1]{Fig.~\ref{#1}}
\newcommand{\Ref}[1]{Ref.~\cite{#1}}
\newcommand{\Sec}[1]{Sec.~\ref{#1}}

\newcommand{\Tab}[1]{Tab.~\ref{#1}}

\newcommand{\Lpb}{\ensuremath{\mathrm{pb}^{-1}}}

\newcommand{\VA}[3]{\ifthenelse{\equal{#2}{#3}}
{\ensuremath{#1\pm#2}}{\ensuremath{#1\,^{+#2}_{-#3}}}}

\newcommand{\PKS}{\ensuremath{K_S}}

\newcommand{\Ppim}{\ensuremath{\pi^-}}

\newcommand{\Ppip}{\ensuremath{\pi^+}}

\renewcommand{\vec}[1]{\mbox{\boldmath{$\rm#1$}}}


\renewcommand{\thesection}{\arabic{section}}%
\makeatletter
\renewcommand{\section}{\@startsection{section}%
{1}%
{0mm}%
{0.35\baselineskip}%%.95
{0.1\baselineskip}%%.5
{\normalfont\large\bf\mathversion{bold}}}%
\makeatother
\makeatletter
\renewcommand{\subsection}{\@startsection{subsection}%
{2}%
{0mm}%
{0.95\baselineskip}
{0.5\baselineskip}
{\normalfont\normalsize\bf\mathversion{bold}}}%
\makeatother

\def\ifm#1{\relax\ifmmode#1\else$#1$\fi}  
\def\DAF{DA\char8NE}
\def\x{\ifm{\times}}  
\def\pt#1,#2,{\ifm{#1\x10^{#2}}}
\def\up#1{\ifm{^{#1}}}  
\def\dn#1{\ifm{_{#1}}}
\def\ab{\ifm{\sim}}

\def\to{\ifm{\rightarrow}} 
\def\kl{\ifm{K_L}}   
\def\ks{\ifm{K_S}}
\def\Kb{\ifm{\rlap{\kern.3em\raise1.9ex\hbox to.6em{\hrulefill}} K}}
\def\ko{\ifm{K^0}}  \def\kob{\ifm{\Kb\vphantom{K}^0}}

\def\pic{\ifm{\pi^+\pi^-}}  

\def\rmk{\rm\kern.5mm }   
\def\f{\ifm{\phi}}  

\def\pb{{\bf p}}

\def\epm{\ifm{e^+e^-}}
\def\sta#1;{\ifm{|\,#1\rangle}}  \def\minus{$-$}  \def\plm{\ifm{\pm}}
\def\figbox#1;#2;{\parbox{#2cm}{%
\vglue3mm\epsfig{file=#1.eps,width=#2cm}\vglue3mm}}

\newcommand{\aff}[2]{Dipartimento di Fisica dell'Universit\`a #1 e Sezione INFN, #2, Italy.}
\newcommand{\affd}[1]{Dipartimento di Fisica dell'Universit\`a e Sezione INFN, #1, Italy.}

\begin{document}
%\maketitle
%\author{KLOE collaboration}
\begin{frontmatter}
\title{\mathversion{bold} First Observation of Quantum Interference in the 
Process \mbox{$\phi \rightarrow$\ks \kl\ $\rightarrow$ \pic \pic}: a Test 
of Quantum Mechanics and $CPT$ Symmetry}
\pagestyle{plain}
\collab{The KLOE Collaboration}

\author[Na]{F.~Ambrosino},
\author[Frascati]{A.~Antonelli},
\author[Frascati]{M.~Antonelli\corauthref{cor1}},
\author[Roma3]{C.~Bacci},
%\author[Roma3]{M.~Barva},
\author[Karlsruhe]{P.~Beltrame},
\author[Frascati]{G.~Bencivenni},
\author[Frascati]{S.~Bertolucci},
\author[Roma1]{C.~Bini},
\author[Frascati]{C.~Bloise},
\author[Roma3]{S.~Bocchetta},
\author[Roma1]{V.~Bocci},
\author[Frascati]{F.~Bossi},
\author[Frascati,Virginia]{D.~Bowring},
\author[Roma3]{P.~Branchini},
%\author[Moscow]{S.~A.~Bulychjov},
\author[Roma1]{R.~Caloi},
\author[Frascati]{P.~Campana},
\author[Frascati]{G.~Capon},
\author[Na]{T.~Capussela},
%\author[Roma2]{G.~Carboni},
\author[Roma3]{F.~Ceradini},
%\author[Pisa]{F.~Cervelli},
\author[Frascati]{S.~Chi},
\author[Na]{G.~Chiefari},
\author[Frascati]{P.~Ciambrone},
\author[Virginia]{S.~Conetti},
\author[Frascati]{E.~De~Lucia},
\author[Roma1]{A.~De~Santis},
\author[Frascati]{P.~De~Simone},
\author[Roma1]{G.~De~Zorzi},
\author[Frascati]{S.~Dell'Agnello},
\author[Karlsruhe]{A.~Denig},
\author[Roma1]{A.~Di~Domenico\corauthref{cor2}},
\author[Na]{C.~Di~Donato},
\author[Pisa]{S.~Di~Falco},
\author[Roma3]{B.~Di~Micco},
\author[Na]{A.~Doria},
\author[Frascati]{M.~Dreucci},
%\author[Roma3]{A.~Farilla},
\author[Frascati]{G.~Felici},
\author[Frascati]{A.~Ferrari},
\author[Frascati]{M.~L.~Ferrer},
\author[Frascati]{G.~Finocchiaro},
\author[Roma1]{S.~Fiore},
\author[Frascati]{C.~Forti},
\author[Roma1]{P.~Franzini},
\author[Frascati]{C.~Gatti},
\author[Roma1]{P.~Gauzzi},
\author[Frascati]{S.~Giovannella},
\author[Lecce]{E.~Gorini},
\author[Roma3]{E.~Graziani},
\author[Pisa]{M.~Incagli},
\author[Karlsruhe]{W.~Kluge},
\author[Moscow]{V.~Kulikov},
\author[Roma1]{F.~Lacava},
\author[Frascati]{G.~Lanfranchi},
\author[Frascati,StonyBrook]{J.~Lee-Franzini},
\author[Karlsruhe]{D.~Leone},
%\author[Frascati,Moscow]{M.~Martemianov},
\author[Frascati]{M.~Martini},
\author[Na]{P.~Massarotti},
%\author[Frascati,Moscow]{M.~Matsyuk},
\author[Frascati]{W.~Mei},
\author[Na]{S.~Meola},
%\author[Roma2]{R.~Messi},
\author[Frascati]{S.~Miscetti},
\author[Frascati]{M.~Moulson},
\author[Frascati]{S.~M\"uller},
\author[Frascati]{F.~Murtas},
\author[Na]{M.~Napolitano},
\author[Roma3]{F.~Nguyen},
\author[Frascati]{M.~Palutan},
\author[Roma1]{E.~Pasqualucci},
%\author[Frascati]{L.~Passalacqua},
\author[Roma3]{A.~Passeri},
\author[Frascati,Energ]{V.~Patera},
\author[Na]{F.~Perfetto},
\author[Roma1]{L.~Pontecorvo},
\author[Lecce]{M.~Primavera},
\author[Frascati]{P.~Santangelo},
\author[Roma2]{E.~Santovetti},
\author[Na]{G.~Saracino},
%\author[StonyBrook]{R.~D.~Schamberger},
\author[Frascati]{B.~Sciascia},
\author[Frascati,Energ]{A.~Sciubba},
\author[Pisa]{F.~Scuri},
\author[Frascati]{I.~Sfiligoi},
\author[Frascati,Novo]{A.~Sibidanov},
\author[Frascati]{T.~Spadaro},
%\author[Roma3]{E.~Spiriti},
%\author[Frascati,Tbilisi]{M.~Tabidze},
\author[Roma1]{M.~Testa\corauthref{cor3}},
\author[Roma3]{L.~Tortora},
\author[Roma1]{P.~Valente},
\author[Karlsruhe]{B.~Valeriani},
\author[Frascati]{G.~Venanzoni},
\author[Roma1]{S.~Veneziano},
\author[Lecce]{A.~Ventura},
%\author[Roma1]{S.Ventura},
\author[Frascati]{R.Versaci},
%\author[Na]{I.~Villella},
\author[Frascati,Beijing]{G.~Xu}
%%%%

%\address[Bari]{\affd{Bari}}
\address[Beijing]{Permanent address: Institute of High Energy 
Physics of Academia Sinica,  Beijing, China.}
\address[Frascati]{Laboratori Nazionali di Frascati dell'INFN, 
Frascati, Italy.}
\address[Karlsruhe]{Institut f\"ur Experimentelle Kernphysik, 
Universit\"at Karlsruhe, Germany.}
\address[Lecce]{\affd{Lecce}}
\address[Moscow]{Permanent address: Institute for Theoretical 
and Experimental Physics, Moscow, Russia.}
\address[Na]{Dipartimento di Scienze Fisiche dell'Universit\`a 
``Federico II'' e Sezione INFN,
Napoli, Italy}
%\address[Novo]{Permanent address: Budker Institute of Nuclear Physics, Novosibirsk, Russia.}
\address[Pisa]{\affd{Pisa}}
\address[Novo]{Permanent address: Budker Institute of Nuclear Physics, Novosibirsk, Russia.}
\address[Energ]{Dipartimento di Energetica dell'Universit\`a 
``La Sapienza'', Roma, Italy.}
\address[Roma1]{\aff{``La Sapienza''}{Roma}}
\address[Roma2]{\aff{``Tor Vergata''}{Roma}}
\address[Roma3]{\aff{``Roma Tre''}{Roma}}
\address[StonyBrook]{Physics Department, State University of New 
York at Stony Brook, USA.}
%\address[Trieste]{\affd{Trieste}}
%\address[Tbilisi]{Permanent address: High Energy Physics Institute, Tbilisi State University, Tbilisi, Georgia.}
\address[Virginia]{Physics Department, University of Virginia, USA.}

\begin{flushleft}
\corauth[cor1]{cor1}{\small $^1$ Corresponding author: Mario Antonelli,
INFN - LNF, Casella postale 13, 00044 Frascati (Roma), 
Italy; tel. +39-06-94032728, e-mail mario.antonelli@lnf.infn.it}
\end{flushleft}
\begin{flushleft}
\corauth[cor2]{cor2}{\small $^2$ Corresponding author: Antonio Di Domenico,
Dipartimento di Fisica dell'Universit\`a ``La Sapienza'' e Sezione INFN, P.le A. Moro 2
00185  Roma,
Italy; tel. +39-06-49914457, e-mail antonio.didomenico@roma1.infn.it}
\end{flushleft}
\begin{flushleft}
\corauth[cor3]{cor3}{\small $^3$ Corresponding author: Marianna Testa
Dipartimento di Fisica dell'Universit\`a ``La Sapienza'' e Sezione INFN, P.le A. Moro 2
00185 Roma, 
Italy; tel. +39-06-49914614, e-mail marianna.testa@roma1.infn.it}
\end{flushleft}

\begin{abstract}

 We present the first observation of quantum interference in the 
 process \f\to\ks\kl\break\to\pic\pic, using the KLOE detector at
 the Frascati \epm\ collider \DAF. From about $5\times 10^4$ neutral kaon pairs both decaying to \pic\ pairs we obtain the distribution of $\Delta t$, the difference
 between the two kaon decay times, which allows testing the validity of quantum mechanics and $CPT$ invariance: no violation of either is observed. New or improved limits on coherence loss and $CPT$ violation are presented.

\end{abstract}
\end{frontmatter}

\let\cl=\centerline
\section{Introduction}
\label{sec:intro}
 A $\phi$-factory provides unique opportunities for testing quantum mechanics 
 (QM) and $CPT$ symmetry. In the decay 
$\phi \rightarrow K^0\bar{K}^0$, 
 the neutral kaon pair is produced in a  $J^{PC}=1^{--}$ state:
\begin{eqnarray}
\sta i;&=&{1\over\sqrt{2}}\:
\left( |K^0,\:\pb\rangle |\bar K^0, -\pb\rangle-|\bar K^0,\:\pb\rangle|K^0,\: -\pb\rangle \right)\nonumber\\
&=& {N\over \sqrt{2}}\:
\left( |K_S,\:\pb\rangle | K_L, -\pb\rangle-|K_L,\:\pb\rangle|K_S,\: -\pb\rangle \right)
\label{eq:state}
\end{eqnarray}  
where \pb\  is  the kaon momentum  in the $\phi$ meson rest frame, and
$N=(1+|\epsilon|^2)/(1-\epsilon^2)$. 
Since $1-N\ll1$ we will set $N=1$ in the following without any loss of generality.
The  decay intensity for the  process  \f\to(2 neutral kaons)\to\pic, \pic\ is then given by \cite{dunietz}:
 \begin{eqnarray}
\label{eq:dtintensity}
I(t_1,\ t_2)&=&\frac{1}{2}\:\left|\,\langle\pic\sta\ks;\,\right|^{\:4}\,\left|\, \eta_{+-}\right|^{\;2}\left(e^{-\Gamma_L t_1 -\Gamma_S t_2}+e^{-\Gamma_S t_1-\Gamma_L t_2}\right. \nonumber\\
& &\kern2cm\left.-2 e^{-(\Gamma_S+\Gamma_L)(t_1+t_2)/2}\cos\,\left(\,\Delta m\,(t_1-t_2)\,\right)\right)
\end{eqnarray}%%\displaystyle
where $t_i$ are the proper  times of the two kaon decays,
$\Gamma_S$ and $\Gamma_L$ are the decay widths of $K_S$ and $K_L$,
$\Delta m = m_L-m_S$ is their mass difference 
and 
$\eta_{+-} = \langle \Ppip\Ppim |K_L\rangle/\langle \Ppip\Ppim |K_S\rangle= |\eta_{+-}|e^{i\phi_{+-}}$.
The two kaons cannot decay into the same final state at the same time, even though the two decays are space-like separated events. Correlations of this type in QM were first pointed out by Einstein, Podolsky, and Rosen (EPR)~\cite{epr}.

While it is not obvious what a deviation from QM might be, the assumption that coherence is lost during the states time evolution does violate QM. One can therefore introduce a decoherence parameter $\zeta$ \cite{eberhard},
simply  multiplying the interference term in Eq.~(\ref{eq:dtintensity}) by a factor of $(1-\zeta)$.
 The meaning and value of $\zeta$ depends on the basis in which the initial state
 (\ref{eq:state}) is written \cite{bertlmann1}.
Eq.~(\ref{eq:dtintensity}) is modified as follows:
\begin{eqnarray}
\label{eq:dtintensitySL}
I(\,t_1,t_2;\zeta_{SL}\,)&=&
\frac{1}{2}\:\left|\,\langle\pic\sta\ks;\,\right|^{\:4}\,\left|\, \eta_{+-}\right|^{\;2}\left(e^{-\Gamma_L t_1 -\Gamma_S t_2}+e^{-\Gamma_S t_1-\Gamma_L t_2}\right. \nonumber\\
& &\kern1cm\left.-2\;(1-\zeta_{SL})\;e^{-(\Gamma_S+\Gamma_L)(t_1+t_2)/2}\cos\,\left(\,\Delta m\,(t_1-t_2)\,\right)\right)
\end{eqnarray}
in the \ks-\kl\ basis, and: 
\begin{eqnarray}
\label{eq:dtintensity00}
I(\,t_1,t_2;\zeta_{0\bar{0}}\,)&=&
\frac{1}{2}\:\left|\,\langle\pic\sta\ks;\,\right|^{\:4}\,\left|\, \eta_{+-}\right|^{\;2}\left(e^{-\Gamma_L t_1 -\Gamma_S t_2}+e^{-\Gamma_S t_1-\Gamma_L t_2}\right. \nonumber\\
&- &2 e^{-(\Gamma_S+\Gamma_L)(t_1+t_2)/2}\cos\,\left(\,\Delta m\,(t_1-t_2)\,\right) \nonumber\\
&+&\frac{\zeta_{0\bar{0}}}{2}\left(-e^{-\Gamma_L t_1 -\Gamma_S t_2}
- e^{-\Gamma_S t_1 -\Gamma_L t_2}\right.\nonumber\\
&+&\left.2 e^{-(\Gamma_S+\Gamma_L)(t_1+t_2)/2}\left(\cos \,\left(\,\Delta m 
(t_1-t_2)\,\right)-\cos\,\left(\,\Delta m (t_1+t_2)\,\right)\right)\right)\nonumber\\
&+&\left.\frac{1}{2}\frac{\zeta_{0\bar{0}}}{|\eta_{+-}|^2}\,e^{-\Gamma_S(t_1+t_2)}\right) \quad \hbox{to the lowest order in $|\eta_{+-}|$ }
\end{eqnarray}
in the $K^0$-$\bar{K}^0$ basis.%

Another phenomenological model~\cite{bertlmann2} introduces decoherence via a dissipative term in the
 Liouville-von Neumann equation for the density matrix of the  state and predicts decoherence to 
 become stronger with increasing
distance between the two kaons. This model introduces a parameter $\lambda$, related to the decoherence parameter in the \ks-\kl\ basis by $\zeta_{SL}\simeq \lambda/ \Gamma_S$.

In a hypothetical quantum gravity, space-time fluctuations
at the Planck scale ($\sim$$10^{-33}$~cm), might induce a pure state to 
become mixed \cite{hawk2}. This results in QM and $CPT$ violation, changing therefore the decay time distribution of the \ko-\kob\ pair from \f\ decays\cite{ellis1}.
Three $CPT$- and QM-violating real parameters,  with dimensions of mass, $\alpha, \beta$, and $\gamma$,  are introduced in \cite{ellis1}. $\alpha, \beta$, and $\gamma$ are guessed to be of $\mathcal {O} (m_{K}^2/M_P) \sim 2 \times 10^{-20}$ GeV \cite{ellis1,ellis2},
 where $M_P=1/\sqrt{G_N}=1.22\times10^{19}$ GeV is the Planck mass.
 The conditions $\alpha=\gamma$ and $\beta=0$ ensure complete positivity  in this framework~\cite{peskin,benatti2,benatti3}.
 The decay intensity is (see Eq. (7.5) in Ref.~\cite{peskin} setting $\alpha=\gamma$ and $\beta=0$):
\begin{eqnarray}\label{eq:peskin}
I(\,t_1,t_2;\gamma\,)&=&
\frac{1}{2}\:\left|\,\langle\pic\sta\ks;\,\right|^{\:4}\,\left|\, \eta_{+-}\right|^{\;2}\nonumber\\
&&\left( \left(1+\frac{\gamma}{\Delta \Gamma |\eta_{+-}|^2}\right)  \left (e^{-\Gamma_L t_1 -\Gamma_S t_2}+e^{-\Gamma_S t_1-\Gamma_L t_2}\right)\right.\\
 &-&\left. 2 \cos\,\left(\,\Delta m\,(t_1-t_2)\,\right) e^{-(\Gamma_S+\Gamma_L)(t_1+t_2)/2}   
  -2\,\frac{\gamma}{\Delta \Gamma |\eta_{+-}|^2} e^{-\Gamma_S(t_1+t_2)}\right)\nonumber
  \end{eqnarray}

 It has been pointed out \cite{mavro1,mavro2} that in this context the initial state (\ref{eq:state}) may acquire a small  $C$-even component: 
\begin{eqnarray}
\sta i;&=&{1\over\sqrt{2}}\:
\left( |K^0,\:\pb\rangle |\bar K^0, -\pb\rangle-|\bar K^0,\:\pb\rangle|K^0,\: -\pb\rangle \right.\nonumber\\
&+&\left. 
\omega \left(|K^0,\:\pb\rangle |\bar K^0, -\pb\rangle+|\bar K^0,\:\pb\rangle|K^0,\: -\pb\rangle
\right)
\right)
\label{eq:state5}
\end{eqnarray}  
where $\omega=|\omega|e^{i\Omega}$ is a complex parameter describing $CPT$ violation,
whose order of magnitude is expected to be at most $|\omega| \sim \sqrt { (m^2_K/M_{P})/\Delta \Gamma } \sim 10^{-3}$, with $\Delta \Gamma = \Gamma_S - \Gamma_L$.
The decay intensity is (see Eq.(3.3) in Ref.~\cite{mavro2} setting $\alpha,\beta,\gamma=0$):
\begin{eqnarray}\label{eq:omega}
I(\,t_1,t_2;\omega\,)&=&
\frac{1}{2}\:\left|\,\langle\pic\sta\ks;\,\right|^{\:4}\,\left|\, \eta_{+-}\right|^{\;2}  \left(
  e^{-\Gamma_S t_1-\Gamma_L t_2} +e^{-\Gamma_L t_1 -\Gamma_S t_2} \right.\nonumber \\
  &-& 2\cos\,\left(\,\Delta \,m\,(t_1-t_2)\,\right) e^{-(\Gamma_S+\Gamma_L)(t_1+t_2)/2} +\frac{|\omega|^2}{|\eta_{+-}|^2} e^{-\Gamma_S(t_1+t_2)}\nonumber \\
&+&2\frac{|\omega|}{|\eta_{+-}|}\left(  \cos\,\left(\, \Delta \, m \,t_2 -\phi_{+-}+\Omega\,\right) e^{-\Gamma_S t_1-(\Gamma_S+\Gamma_L)t_2/2}  \right. \nonumber \\
&-&\left. \left. \cos\,\left(\, \Delta \, m \,t_1-\phi_{+-}+\Omega\,\right) e^{-\Gamma_S t_2-(\Gamma_S+\Gamma_L)t_1/2} \right)\right)
 \end{eqnarray}
The decoherence parameters $\zeta_{SL}$ and $\zeta_{0\bar{0}}$, have been found in the past to be compatible with zero, with uncertainties of 0.16 and 0.7, respectively, using CPLEAR data~\cite{bertlmann1,bertlmann2,cplear1}.
CPLEAR has also analyzed single neutral-kaon decays to
measure the
$\alpha$, $\beta$, and $\gamma$ parameters \cite{cplear:abg}.
The values obtained for all three parameters are compatible with
zero, with uncertainties of $2.8 \times 10^{-17}\mbox{~GeV}$,
 $2.3 \times 10^{-19}\mbox{~GeV}$, and $2.5 \times 10^{-21}\mbox{~GeV}$,
respectively.
The parameter $\omega$ has never been measured.

In the following, the improved KLOE measurements of the $\zeta_{SL}$, $\zeta_{0\bar{0}}$, $\lambda$, $\gamma$, and $\omega$ parameters are presented. The analysis is based on data collected at \DAF\ in 2001--2002, corresponding to an integrated luminosity of $L\approx 380$~\Lpb. \DAF, the Frascati \f\ factory, is an $e^+e^-$ collider operated at a center of mass energy $W=M$(\f)\ab1020 MeV. Electrons and positrons collide in the horizontal plane at an angle of $\pi$\minus25 mrad. \f-mesons are produced with approximately 12 MeV/c momentum toward the rings center, along the $x$-axis. The $z$-axis is taken as the bisectrix of the two beams, the $y$-axis being vertical.

\section{The KLOE detector}
The KLOE detector consists of a large, cylindrical drift chamber (DC),
surrounded by a 
lead/scintillating-fiber electromagnetic calorimeter (EMC).
 A superconducting coil around the calorimeter 
provides a 0.52 T field. 
The drift chamber \cite{KLOE:DC} is 4~m in diameter and 3.3~m in length.
The momentum resolution is $\sigma_{p_{\perp}}/p_{\perp}\approx 0.4\%$. 
Two-track vertices are reconstructed with a spatial resolution of $\sim$ 3~mm. 
The calorimeter \cite{KLOE:EmC} is divided into a barrel and two endcaps.
 It covers 98\% of the solid angle.
 Cells close in time and space are grouped into calorimeter clusters.
 The energy and time resolutions for photons of energy $E$ are
 $\sigma_E/E = 5.7\%/\sqrt{E\ {\rm(GeV)}}$
 and $\sigma_t = 57\ {\rm ps}/\sqrt{E\ {\rm(GeV)}}\oplus100\ {\rm ps}$, respectively.
The KLOE trigger \cite{KLOE:trig} uses calorimeter and chamber information. For this 
analysis, only the calorimeter signals are used. Two energy deposits above threshold
 ($E>50$ MeV for the barrel and  $E>150$ MeV for the endcaps) are required.

Kaon regeneration in the beam pipe is a non negligible disturbance.
The beam pipe is spherical around the interaction point, with 
a radius of 10 cm. The walls of the beam pipe, 500 $\mu$m thick, are made of a
62\%-beryllium/38\%-aluminum alloy (AlBeMet$\circledR$162).
A beryllium cylindrical tube of 4.4 cm radius and 50$\mu$m thick, coaxial with the beam, provides electrical continuity. 

We only use runs satisfying basic quality criteria.
For each run we determine the average collision conditions: \pb=\pb\dn{e^+}+\pb\dn{e^-}=\pb\dn{\phi}, the center of mass energy W, the beam bunch dimensions, the collision point $\vec{r}$\dn{\rm C} and angle. This is done using Bhabha scattering events.
We then require $|p_{y,z}|<$ 3 MeV, $|W -1020|<$ 5 MeV.
The collision point must satisfy $|\,x_{\rm C}\,|<$ 3 cm, $|\,y_{\rm C}\,|<$ 3 cm, and $|z_{\rm C}|<$ 5 cm. The rms spread of the luminous region must satisfy $\sigma_x<$ 3 cm and $\sigma_z<$ 3 cm. A small number of runs were rejected because of improper trigger operation.
Each run used in the analysis is simulated with the KLOE Monte Carlo (MC) simulation program, GEANFI \cite{KLOE:offline}, using values of relevant machine 
parameters such as $W$ and \pb\dn{\f} determined as mentioned above.
Machine background obtained from data is superimposed on MC events
on a run-by-run basis.
For \ks\kl\to\pic\pic\ events, the number of simulated events is 10
times that expected on the basis of the integrated luminosity.
For all other processes, the effective statistics of the simulated sample
and of the data sample are approximately equal.  
The effects of initial- and final-state radiation are included in the
simulation. Final-state radiation in \ks\ and \kl\ decays is treated as
discussed in Ref.~\cite{KLOE:gattip}.

\section{Analysis}

\subsection{Event selection}

Events are selected on the basis of two identified neutral kaons from $\phi$ decay, in turn decaying 
into \Ppip\Ppim\ pairs.
Since we cannot tell whether \pic\ decays near the interaction point (IP) are from a \ks\ or a \kl, we try to ensure that the same criteria are used for all decays.
In the following we call $K_1$ ($K_2$) the 
decay closest to (farthest from) the $\phi$ production point.  

 We first require a $K_1$ vertex with two tracks of opposite curvature
 within a fiducial volume with $r<10$ cm and $|\,z\,|<20$ cm centered
 at the nominal collision point $\vec{r}$\dn{\rm C},
 determined as discussed above.
 We also require that the two tracks satisfy $|\,m_{\pic}-m_{K}\,|<5$ MeV and $|\,p_{\pic}-p_K\,|<$ 10 MeV/$c$, where $p_K$  is calculated from the kinematics of \f\to\ks\kl. $m_{\pic}$ and  $p_{\pic}$ are respectively the  invariant mass and the momentum of the \pic\ pair.

 In order to search for a second \pic\ kaon decay ($K_2$), 
 all relevant tracks in the chamber---after removal of those 
originating from the decay already identified---are extrapolated 
to their points of closest approach to the $K_2$ path computed from kinematics
 and $\vec{x}$\dn{\rm C}.
For each track candidate we compute $d$, the distance of closest approach to the $K_2$ path. 
For each charge we take the tracks with smallest value of $d$ as the $K_2$ decay pions. 
We then determine the two track vertex and require that $|\,m_{\pic}-m_{K}\,|<5$ MeV 
and $|\,p_{\pic}-p_K\,|<$ 10 MeV/$c$. 
 We also require \minus50$<E^{\,2}_{\rm miss}- p_{\rm miss}^{\,2}<10$  MeV\up{\,^2} and
 $\sqrt{E^{\,2}_{\rm miss}+p_{\rm miss}^{\,2}}<10$ MeV, where 
 $p$\dn{\rm miss} and $E_{\rm miss}$ are the missing momentum and energy computed assuming
 $K_2$\to\pic.

\begin{figure}[ht]
  \begin{center}
    \resizebox{0.6\textwidth}{!}{\includegraphics{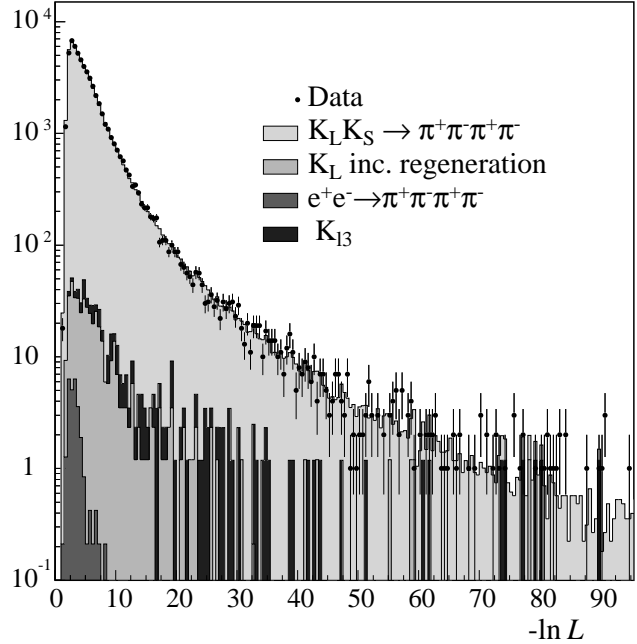}}
  \end{center}
  \caption
      {Distribution of the minimum of $-\ln  {\it L}$ of the
 kinematic fit for data and Monte Carlo for different decay channels.}
\label{fig:chi2_data_mc}
\end{figure}

 More accurate values for the $K_{1,2}$ vertex positions, $\bar{\vec{r}}_{1,2}$ and 
the collision point, $\bar{\vec{r}}\dn{\rm C}$, are obtained from a kinematical fit.
The fit makes use of the 
constraint  from the $K_1$ and $K_2$ directions $\vec{n}_{1,2}$ defined by:
\[ \vec{n}_{1,2} = \frac {(\vec p_++\vec p_-)_{1,2}-(\vec p_++\vec p_- )_{2,1}+\vec p_\f}{|(\vec p_++\vec p_-)_{1,2}-(\vec p_++\vec p_- )_{2,1}+\vec p_\f|}. \]
We then have
$$\bar{\vec{r}}_{i} = \bar{\vec{r}}_{\rm C} + l_{i}\vec{n}_{i},\qquad  i=1,\ 2$$
and solve for $\bar{\vec{r}}_{\rm C}$ and $l_{1,2}$ by maximizing the likelihood
  \[ \ln  {\it L} = \sum_{i=1,2} \ln P_i(\vec{r}_i, \bar{\vec{r}}_{\rm C} + l_{i}\vec{n}_{i}) + \ln P_{\rm C}(\vec{r}_{\rm C},\bar{\vec{r}}_{\rm C}) , \] 
 where $P_i$  and $P_{\rm C}$  are the probability density functions for 
 $\vec{r}_{i}$ and $\vec{r}_{\rm C}$, respectively,
 as obtained from MC. The value of $l_{1,2}$ has been kept  positive in the maximization. 
 The minimum of $- \ln  {\it L}$ is shown in \Fig{fig:chi2_data_mc} for data and MC.
  In order to maximize signal efficiency, improve $\Delta t$
resolution, and minimize incoherent regeneration background, as a final
selection requirement we retain events with  $ -\ln  {\it  L }< 7.5$.

 The time difference between $K_1\to \Ppip\Ppim$  and $K_2\to \Ppip\Ppim$ 
is determined as $\Delta t=|t_{1}-t_{2}|$, where  the proper time
is  $t_{i}=l_i/(\beta_{i}\gamma_{i})$
with $\beta_{i}\gamma_{i} = p_{K_i}/m_{K}$.

\subsection{Determination of  background}\label{sec:irr_bkg}

 The selected $\ks\kl\to\pic\pic$ events have a contamination of \ab3.2\% for $\Delta t<35\tau_S$
 dominated by regeneration on the beam pipe.
 Semileptonic \kl\ decays amount to $0.2\%$ as determined from MC.
 Direct four pion production, $e^+e^- \to \pi^+\pi^-\pi^+\pi^-$,  gives a \ab0.3\%  contamination
 at the IP with $\Delta t\sim 0$, the region most sensitive to coherence loss. This contribution
 is obtained from the sidebands: $10\hbox{ MeV}<\sqrt{E^2_{\rm miss}+p_{\rm miss}^2}<20 \hbox{ MeV}$
  and  $|p_{\pic}-p_{K}|>1$ MeV.
Vertex positions, total energy $E_{4\pi}$, and total momentum $p_{4\pi}$ are used to 
distinguish between \epm\to\pic\pic\ events and semileptonic \kl\ decays near the IP.
We find 27\plm8 events, in agreement with the estimate of \ab32 events from the cross section given in  \Ref{cmd2:4pi}.

Incoherent (see Sec.~\ref{sec:fit}) and coherent regeneration on the beam-pipe are included in the fit of the $\Delta t$ 
distribution. The coherent regeneration amplitude, $\rho_{\rm coh}=|\rho_{\rm coh}|e^{i\phi_{\rm coh}}$,
is obtained from the time distribution
for $\kl(\to \ks) \to \pic$ decays after single \kl's cross the beam pipe: 
\begin{eqnarray}
\label{eq:prop_dec}
I(t) &=&\left|\langle \pic\sta K_{L}(t)+\rho_{\rm coh}|K_{S}(t);\right|^2  \\
  &=&\left|\langle \pic|\ks\rangle\right|^2 \left(|\eta_{+-}|^2\,e^{-\Gamma_Lt}+ |\rho_{\rm coh}|^2e^{-\Gamma_S t} 
\right.\nonumber\\  &+&2 \left.|\eta_{+-}|\, |\rho_{\rm coh}| \,e^{-(\Gamma_S+\Gamma_L)t/2}\cos(\Delta m t+\phi_{\rm coh}-\phi_{+-})\right)\nonumber
\end{eqnarray}  
\kl's are identified by the reconstruction of a \ks\to\pic\ decay,
 using the same algorithm used for the measurement
 of the $\kl\to \pic$ branching ratio \cite{KLOE:brpp}.
 In addition we require $|\,p_{K_L}-p_{\pic}\,|<5$ MeV
 and $|E_{\pic} -\sqrt{p_{K_L}^2+m_{K}^2}+m_{\pic}-m_{K}|<5$ MeV,
 where the \kl\ momentum, $p\dn{\kl}$, is obtained from the \ks\ direction
 and \pb\dn\f\ and 
$E_{\pic}$ is the  energy of the \pic\ pair.
 These cuts are effective for the identification of
 \kl\to\pic\ decays and coherent and incoherent  \kl\to\ks \to \pic regeneration
 processes with a negligible amount of background.
 Incoherent regeneration is rejected by requiring that the angle between \pb\dn{\kl} and \pb\dn{\pic} be smaller than 0.04. 
The residual
 contamination is \ab3\%, from the sidebands
 of the distribution in the above angle.
  Fitting the proper-decay-time distribution, Eq.~(\ref{eq:prop_dec}), 
  we obtain $|\rho_{\rm coh}| = (6.5 \pm 2.2)\times 10^{-4}$ and
  $\phi_{\rm coh} = (-1.05 \pm 0.25)$ rad.
  This result is stable against variations of the scattering angle cut and agrees with  predictions~\cite{rege}.
Coherent regeneration in the inner pipe is negligible.
Background due to production of $C$-even neutral kaon pairs in two photon processes or $(f_0,\ a_0)$ decays is also negligible \cite{oller,achasov1,achasov2,dunietz}.

\subsection{Determination of the detection efficiency}
\begin{figure}
  \begin{center}
    \resizebox{0.6\textwidth}{!}{\includegraphics{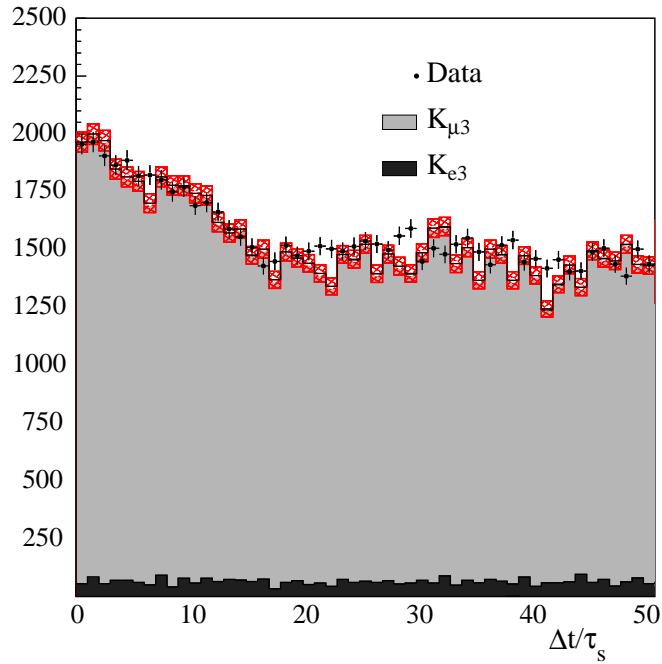}} 
  \end{center}
  \caption{ $\Delta t$ distribution for the
 $K_SK_L \rightarrow \pi^+ \pi^-, \pi \mu \nu$  control sample for data (black points)
 and Monte Carlo (solid histogram). The expected background contamination from  $K_SK_L \rightarrow \pi^+ \pi^-, \pi e \nu$ is also shown. The  hatched area represents the  Monte Carlo statistical uncertainty.}
  \label{fig:dt-mc_corrdist}
\end{figure}

 The overall detection efficiency is about 30\%, and
 has contributions from the event reconstruction and
 event selection efficiencies. 
 These efficiencies have been evaluated from MC. 
 For the reconstruction efficiency, a correction obtained from data
 is applied. This correction is determined
 using an independent sample of $K_SK_L \rightarrow \pi^+\pi^-, \pi \mu \nu$
 decays. 
 $\pi \mu \nu$  decays are identified by requiring 
 $\sqrt{E^2_{miss}+ p_{\rm miss}^2}>$ 10 MeV,  $|p^*_{\pm}|<246$ MeV and
 $p^*_{+} + p^*_{-}<367$ MeV,   where $p^*$  is the momentum of the
 decay secondary in the kaon rest-frame, calculated assuming  the $\pi$ mass
 hypothesis. We then compute the squared lepton mass  $m^2_{l^-}$($m^2_{l^+}$)
 in the hypothesis $K \to  \pi^+ \mu^- \bar{\nu}$ ($K \to  \pi^- \mu^+ \nu$),
  and require:
  $150\, {\rm MeV}^2 < m^2_{l^-}+m^2_{l^+}<270\,{\rm MeV}^2$. 
The distribution of the time difference $\Delta t=|t_1-t_2|$ between the
 two kaon decays obtained for the
 $K_SK_L \rightarrow \pi^+ \pi^-, \pi \mu \nu$ sample is
 shown in \Fig{fig:dt-mc_corrdist}, both for data and MC.
The correction to the reconstruction efficiency from the MC
is obtained from the data-MC ratio of the distributions in
 \Fig{fig:dt-mc_corrdist} and
 applied bin by bin as a function of $\Delta t$.

In order to take into account the $\Delta t$ resolution when fitting,
a smearing matrix has been constructed from MC by 
filling a two-dimensional histogram with the ``true'' and reconstructed
values of $\Delta t$. The efficiency correction and the smearing matrix 
are then used in the fit procedure as explained in the following section.  

\subsection{Fit}\label{sec:fit}
\begin{figure}[t!]
   \begin{center}
    \resizebox{0.6\textwidth}{!}{\includegraphics{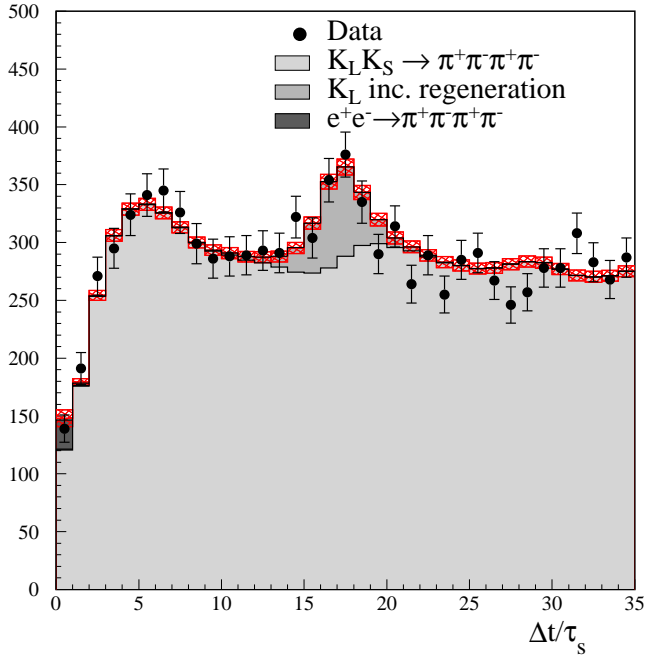}}
   \end{center}
   \caption
       { $\Delta t$ distribution from the fit used to determine 
$\zeta_{SL}$. The black points with errors are data and the solid histogram
 is the fit result.
 The uncertainty arising from the efficiency correction is shown as the
  hatched area.}
       \label{fig:fit_dec_def}
 \end{figure}

We fit the observed $\Delta t$ distribution between 0 and 35$\tau_S$ in 
intervals of $\overline{\Delta t}= \tau_S$. The fitting function is obtained 
from the  $I(t_1,t_2)$ distribution given in Eq.~(\ref{eq:dtintensity})
including the QM  violating parameter $\zeta$, or  the QM 
and $CPT$ violating parameters $\gamma$  and $\omega$  as discussed in Sec.\ref{sec:intro}.
To take coherent regeneration into account in Eq.~(\ref{eq:dtintensity}),
  the time evolution of the single kaon is modified as follows: %if the decay time $t_i$ is greater than the time needed to reach one regenerator:
\begin{eqnarray}\label{eq:coh_reg}
  |K_{S,L}(t_i) \rangle   =   |K_{S,L}(t_i) \rangle  + \rho_{\rm coh} |K_{L,S}(t_i)\rangle
\label{eq:regen}
\end{eqnarray}  
where $\rho_{\rm coh}$
is evaluated as explained above. 
We then integrate $I(t_1,t_2)$ over the sum $t_1+t_2$ for fixed $\Delta t=|t_1-t_2|$, and over the bin-width of the data histogram:
$$I_j(\vec{q})=\int_{(j-1) \bar{\Delta t}}^{j \bar{\Delta t}}\!\!  d(\Delta t)  \int_{\Delta t}^{\infty}  {I(t_1,t_2;\vec{q})}\:d(t_1+t_2),$$
where $\vec{q}$ is the vector of the QM- and $CPT$-violating parameters.
 Finally, the observed $\Delta t$ distribution is fitted with the following function:
 \begin {equation}
   n_i = N\left( \sum_j s_{ij}\:\epsilon_j\:I_j(\vec{q})\right)+N^{\rm reg} I^{\rm reg}_i + N_{4\pi}{I_{4\pi ,i}} 
% I^{sph\,pipe}_j + I^{cyl\,  pipe}_j)    
 \label{eq:fit_eq}
 \end {equation}
 where $n_i$ is the expected number of events in the $i^{\rm th}$ bin of the histogram, $s_{ij}$
 is the smearing matrix, and $\epsilon_j$ is the efficiency.  $N$, the number of $K_SK_L\to \pic\pic$ events, and  $N^{\rm reg}$, the number of events due to incoherent regeneration, are free parameters in the fit. 
The time distribution $I^{\rm reg}_i$ for the contribution from incoherent regeneration is evaluated  from MC.
The contribution from non-resonant $e^+e^-\to\pic\pic$ events is treated
in a similar manner, except for that $N_{4\pi}$ is fixed to the value 
determined as in \Sec{sec:irr_bkg}, rather than left free in the fit.
The fit is performed by minimizing the least squares function:
 \begin{eqnarray}
   \chi^2 &=& \sum_{i=1}^{n}{(N_i^{\rm data}-n_i)^2/\left(n_i + 
(n_i\delta \epsilon_i/\epsilon_i)^2\right)}
 \end{eqnarray}
 where $N_i^{\rm data}$ is the number of events observed in the $i^{\rm th}$
bin and %$n_i$ are the number of  expected events,
$\delta \epsilon_i$ is the error on the efficiency, including
 the correction. Using Eq.~(\ref{eq:fit_eq}) with the QM- and $CPT$-violating parameters fixed to zero, $\Delta m$  can be left
 as free parameter and evaluated.
 In this case, the fit gives 
$$\Delta m = (5.61 \pm 0.33)\times 10^{9} \, \mbox{s}^{-1}, $$ which
 is compatible with the more precise value given by the PDG~\cite{PDG2006}: $$\Delta m = (5.290 \pm 0.015)\times 10^{9}\,  \mbox{s}^{-1}.$$
For the determination of the QM- and $CPT$-violating parameters,
$\Delta m$ is fixed to the PDG value in all subsequent fits.
 As an example, the fit of the $\Delta t$ distribution used to determine 
 $\zeta_{SL}$
 is shown in \Fig{fig:fit_dec_def}: the peak in the vicinity of 
 $\Delta t \sim 17\, \tau_{S}$ is due to coherent and incoherent
 regeneration on the spherical beam pipe.

\subsection {Systematic uncertainties}
As possible contributions to the systematic uncertainties on the 
QM- and $CPT$-violating parameters determined, we have considered 
the effects of data-MC discrepancies (in particular on the $\Delta t$
resolution), dependences on cut values, and imperfect knowledge of
backgrounds and other input parameters.
The contributions from each source to the systematic uncertainty 
on each parameter determined are summarized in \Tab{tab:syst}, 
and discussed in further detail in the following.
\begin{table}[htb]
\caption{Summary of systematic uncertainties}
\label{tab:syst}
\newcommand{\cc}[1]{\multicolumn{1}{c}{#1}}
\renewcommand{\tabcolsep}{0.5pc}
 \renewcommand{\tabcolsep}{0.2pc} % enlarge column spacing
\renewcommand{\arraystretch}{1.2} % enlarge line spacing
\begin{tabular}{@{}llllll}
\hline
 &Cut &    Resolution&Inputs&Coherent&$\pic\pic$  \\
\noalign{\vglue-7mm}\\
&stability &         &       &reg.        &bckgnd\\
\hline
$\delta \,\zeta_{SL}$ & 0.007& 0.002& 0.001&0.001&0.020\\
$\delta \,\zeta_{0\bar{0}}\times 10^{5}$ & 0.03& 0.01& 0.01 &-&0.11\\
$\delta \,\gamma \times 10^{21}$ GeV &0.4 &0.2 &0.1 &0.3&1.3\\
$\delta$ $\Re\,\omega \times 10^{4}$ & 0.8 &  0.1&  0.1& 0.3&1.4\\
$\delta$ $\Im\,\omega\times 10^{4}$ & 0.4 &  0.4&  0.2&0.3&0.1\\
\hline
\end{tabular}
\end{table}
 Since the QM- and $CPT$- violating parameters are most sensitive to small
values of $\Delta t$, particular attention has been devoted to the evaluation 
of systematic effects in that region.
 The dependence of the detection efficiency on $\Delta t$ is mostly due to 
 the cut on $-\ln {\it L}$ in the kinematic fit used to evaluate the 
 vertex positions.  We have varied the cut from 6.5 to 8.5, corresponding to a fractional
 variation  in the efficiency of $\sim 5\%$.
 The corresponding changes in the final results are consistent with statistical
 fluctuations.
 For each physical parameter determined, we take the systematic error 
 from this source to be half of the difference
 between the highest and lowest parameter values obtained as a result of
this study. These contributions are listed in the first column of 
\Tab{tab:syst}.

\begin{figure}[hb]
   \begin{center}
     \resizebox{0.6\textwidth}{!}{\includegraphics{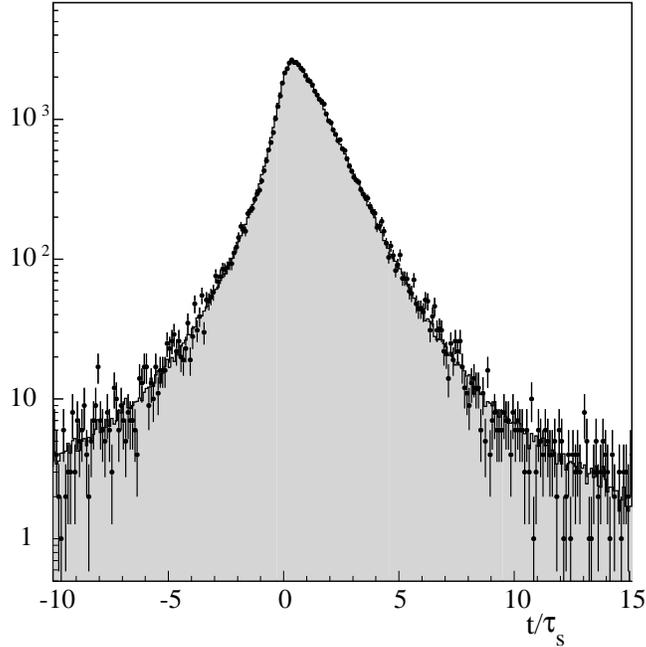}}
   \end{center}
   \caption
 {Distribution of \PKS\ proper-time distribution for data (black points) and Monte Carlo
 (solid histogram).}
      \label{fig:kslt_dtmc}
\end{figure}

 The QM- and $CPT$-violating  parameters depend also on the $\Delta t$
 resolution. We have checked the reliability of the MC simulation on a sample of events with 
 $K_1$=\ks\ $K_2$=\kl\ by requiring $l_2>12$ cm and removing the $l_1>0$ cut.
 \Fig{fig:kslt_dtmc} shows the \ks\ proper-time distribution for data and
 MC. From the negative tail of the \ks\ proper-time distribution we obtain the experimental resolution.
We fit the  data and MC distributions  to an exponential function 
convoluted with the resolution.  We obtain an rms spread of
$ (1.152 \pm 0.020)\,\tau_S$ for data and  $ (1.1807 \pm  0.0036)\,\tau_S$ for MC, i.e.,
the data and MC resolutions agree to within 1.4 standard deviations. 
In addition, we obtain a \PKS\ lifetime, 
$\tau_{S} = (0.9030 \pm 0.056)\,10^{-10}$~s,
in agreement with the world average value 
$(0.8958 \pm 0.0005)\,10^{-10}$~s \cite{PDG2006}.
To estimate the resulting contributions to the systematic uncertainties
on the QM- and $CPT$-violating parameters, the MC resolution is varied by 
$\pm$5\%, about three times the statistical uncertainty of the check. 
For each parameter determined, we take the systematic error from this 
 source to be half of the difference
 between the highest and lowest values obtained.
 These contributions are listed in the second column of 
\Tab{tab:syst}.
The third column gives the contributions of uncertainties on the known values of $\Delta m$, $\Gamma_L$, $\Gamma_S$ 
 and $\eta_{+-}$, which  have been
 propagated numerically.
%Their contributions are listed in the third
%column of \Tab{tab:syst}.

Contributions to the systematic uncertainties due to limited knowledge 
on $|\rho_{\rm coh}|$ and $\phi_{\rm coh}$  
 have been evaluated by varying the parameter values within their errors and are listed in the  fourth column of  \Tab{tab:syst}.
% The systematic uncertainties are evaluated by summing the contributions
% from each parameter in quadrature, and listed in the fourth column of 
% \Tab{tab:syst}.
 
Finally the last column gives the  contributions arising from  the uncertainty on the level of background contamination from non-resonant $e^+e^-\to \pi^+\pi^-\pi^+\pi^-$ events, evaluated  by varying the background parameter in the fits
 ($N_{4\pi}$ in \Eq{eq:fit_eq}) within its error.
% Finally, the uncertainties on the parameters determined
% arising from the uncertainty in the level of background contamination
% from non-resonant $e^+e^-\to \pi^+\pi^-\pi^+\pi^-$ events
% are also obtained by varying the background parameter in the fits
% ($N_{4\pi}$ in \Eq{eq:fit_eq}) within its error range.
% The corresponding contributions to the uncertainties on each parameter
% are listed in the last column of \Tab{tab:syst}. 
 Note, however, 
 that these contributions are included in the statistical uncertainties, rather than in the 
 systematic uncertainties, in the statement of the final results.

\section{Results and conclusions}

From the fit we obtain the decoherence parameter values:
\begin{eqnarray*}
\zeta_{SL}&=&0.018\,\pm 0.040_{\mbox{stat}}\pm 0.007_{\mbox{syst}}  \qquad \qquad ~~ \chi^2/\mbox{dof}  =  29.7/32;\\
\zeta_{0\bar{0}}&=&\left(0.10\, \pm 0.21_{\mbox{stat}}\pm 0.04_{\mbox{syst}}\right)\times 10^{-5} \qquad \chi^2/\mbox{dof}  =  29.6/32.
\end{eqnarray*}
 which are consistent with $\zeta = 0$ and no QM modification.
 Using the Neyman procedure\cite{feldmancousins}, we derive the upper limits
 $\zeta_{SL} < 0.098$ and
 $\zeta_{0\bar{0}} < 0.50 \times 10^{-5}$ at 95\% C.L.
 Since decoherence in the $K^0\bar{K}^0$ basis would result in the $CP$ allowed $K_SK_S\to\pi^+\pi^-\pi^+\pi^-$ decays, Eq. \ref{eq:dtintensity00}, the value for
 $\zeta_{0\bar{0}}$ is naturally much smaller. 
 In the model of \Ref{bertlmann2},  we find:
$$\lambda=(0.13 \pm 0.30_{\mbox{stat}}\pm 0.05_{\mbox{syst}} )\times 10^{-15}\ \mbox{GeV},
\quad\lambda<0.73 \times 10^{-15}\ \mbox{GeV at 95\% C.L.}.$$
All the above results are a considerable improvement on those obtained from CPLEAR data \cite{bertlmann1,bertlmann2}.
We have measured the $\gamma$ parameter:
$$\gamma = \left(1.3^{+2.8}_{-2.4}\pm 0.4\right)\times 10^{-21} \,\mbox{GeV}$$
with $\chi^2/\mbox{dof} = 33/32$. From the above we find  
 $\gamma < 6.4\times 10^{-21} \,\mbox{GeV}$ 
 at 95 \% C.L.. 
% $\gamma < 5.5\times 10^{-21} \,\mbox{GeV}$ 
% at 90 \% C.L. 
This result is competitive with that obtained by CPLEAR~\cite{cplear:abg}
 using single kaon beams.
\begin{figure}[h!]
   \begin{center}
     \resizebox{0.6\textwidth}{!}{\includegraphics{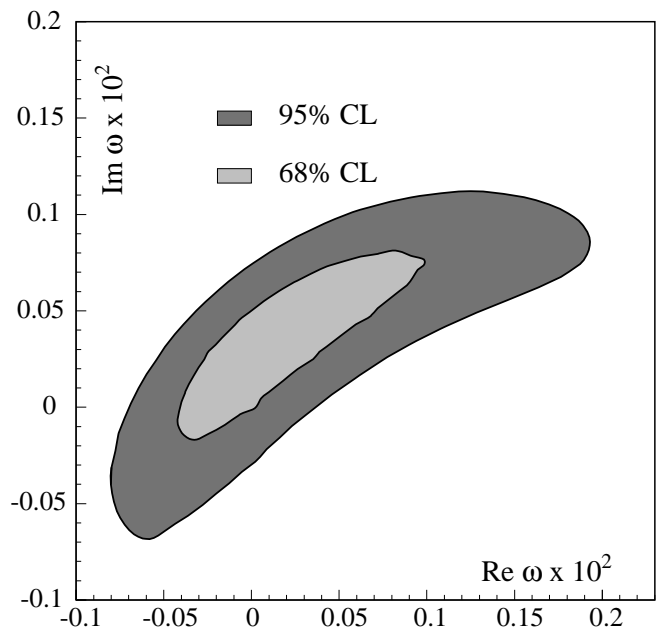}}
   \caption
       {Contour plot of $\Im\,\omega$ versus $\Re\,\omega$ at 
68\% and 95\% C.L.}
       \label{fig:reomvsimom}
\end{center}
 \end{figure}\\
The complex parameter $\omega$ has been measured for the first time. 
The result is 
$$\Re\,\omega =\left( 1.1^{+8.7}_{-5.3} \pm 0.9 \right)\times{10^{-4}}\qquad
\Im\,\omega =\left( 3.4^{+4.8}_{-5.0} \pm 0.6 \right)\times{10^{-4}}$$
with  $\chi^2/\mbox{dof} = 29/31$. The correlation coefficient between  ${\Re\,\omega}$ and   ${\Im\,\omega}$ is 90\%. 
\Fig{fig:reomvsimom} gives the 68\%  and 95\% C.L. contours in the $\Im\,\omega,\ \Re\,\omega$  plane. The upper limit is $|\omega|<2.1 \times 10^{-3}$ at 95\% C.L.\\

We do not find any evidence for QM or $CPT$ violation.\\

\section*{Acknowledgments}
We thank the DA$\Phi$NE team for their efforts in maintaining low background running 
conditions and their collaboration during all data-taking. We want to thank our technical staff: 
G.F.Fortugno for his dedicated work to ensure an efficient operation of the KLOE Computing Center; 
M.Anelli for his continuous support to the gas system and the safety of the detector; 
A.Balla, M.Gatta, G.Corradi and G.Papalino for the maintenance of the electronics;
M.Santoni, G.Paoluzzi and R.Rosellini for the general support to the detector; 
C.Piscitelli for his help during major maintenance periods.
One of us, A.D.D., wishes to thank J.~Bernabeu, R.~A.~Bertlmann, J.~Ellis, M.~Fidecaro, R.~Floreanini, B.~C.~Hiesmayr, and N.~Mavromatos, for private discussions.
This work was supported in part by DOE grant DE-FG-02-97ER41027; 
by EURODAPHNE, contract FMRX-CT98-0169; 
by the German Federal Ministry of Education and Research (BMBF) contract 06-KA-957; 
by Graduiertenkolleg `H.E. Phys. and Part. Astrophys.' of Deutsche Forschungsgemeinschaft,
Contract No. GK 742; by INTAS, contracts 96-624, 99-37; 
by TARI, contract HPRI-CT-1999-00088.

\bibliographystyle{elsart-num}
\bibliography{kl-ks_add2}

\begin{thebibliography}{10}
\expandafter\ifx\csname url\endcsname\relax
  \def\url#1{\texttt{#1}}\fi
\expandafter\ifx\csname urlprefix\endcsname\relax\def\urlprefix{URL }\fi

\bibitem{dunietz}I.~Dunietz, J.~Hauser, J.~L.~Rosner, \Journal{\PRD}{35}{2166}{1987} 
%%\bibitem{buch} C.D.~Buchanan {\it et al.},  \Journal{\PRD} {45}{4088}{1992} 
\bibitem{epr} A.~Einstein, B.~Podolsky, N.~Rosen, \Journal{\PR} {47}{777}{1934}
\bibitem{eberhard} P.H.~Eberhard, ``Tests of Quantum Mechanics at a \f\ factory'' in {\it The second Da$\phi$ne handbook,  Vol.I},
  ed. L.~Maiani, G.~Pancheri, N.~Paver, (1995) 99
%\bibitem{furry} W.~H.~Furry, \Journal{\PRD} {49} {393} {1936}
\bibitem{bertlmann1} R.~A.~Bertlmann, W.~Grimus, B.~C.~Hiesmayr,  \Journal{\PRD} {60} {114032} {1999}
\bibitem{bertlmann2} R.~A.~Bertlmann, K.~Durstberger, B.~C.~Hiesmayr, 
 \Journal{\PRA} {68}{012111}{2003}
\bibitem{hawk2} S.~Hawking, {\it Commun. Math. Phys.}  {\bf 87} (1982) 395
\bibitem{ellis1} J.~Ellis, J.~S.~Hagelin, D.~V.~Nanopoulos, M.~Srednicki, \Journal{\NPB } {241} {381} {1984}
\bibitem{ellis2} J.~Ellis, J.~L.~Lopez, N.~.E.~Mavromatos, D.~V.~Nanopoulos, \Journal{\PRD}{53}{3846}{1996}
\bibitem{peskin} P.~Huet, M.~Peskin,  \Journal{\NPB}{434}{3}{1995}
\bibitem{benatti2} F.~Benatti, R.~Floreanini, \Journal{\NPB} {511}{550}{1998}
\bibitem{benatti3} F.~Benatti, R.~Floreanini, \Journal{\PLB} {468}{287}{1999}
\bibitem{mavro1} J.~Bernabeu, N.~Mavromatos, J.~Papavassiliou, \Journal{\PRL} {92}{131601}{2004}
\bibitem{mavro2} J.~Bernabeu, N.~Mavromatos, J.~Papavassiliou, A.~Waldron-Lauda, \Journal{\NPB}{744}{180}{2006}
\bibitem{cplear1} A.~Apostolakis {\it et al.}, CPLEAR Collaboration, \Journal{\PLB} {422} {339} {1998}
\bibitem{cplear:abg} R.~Adler {\it et al.}, CPLEAR Collaboration, \Journal{\PLB}{364}{239}{1995}


\bibitem{KLOE:DC}
M.~Adinolfi {\it et~al.}, KLOE Collaboration, \Journal{\NIMA} {488}{51}{2002}


\bibitem{KLOE:EmC}
M.~Adinolfi {\it et~al.}, KLOE Collaboration, \Journal{\NIMA} {482}{364}{2002}

\bibitem{KLOE:trig}
M.~Adinolfi {\it et~al.}, KLOE Collaboration, \Journal{\NIMA} {492}{134}{2002} 


%\bibitem{KLOE:brl}
% F.~Ambrosino {\it et~al.}, KLOE Collaboration \Journal{\PLB} {632} {43} {2006}

\bibitem{KLOE:offline}
F.~Ambrosino {\it et~al.}, KLOE Collaboration,  \Journal{\NIMA} {534}{403}{2004}

\bibitem{KLOE:gattip}
C.~Gatti, \Journal{\EPJ} {45}{417}{2006}

\bibitem{cmd2:4pi} R.R. Akhmetshin {\it et~al.}, \Journal{\PLB} {595}{101}{2004}
\bibitem{KLOE:brpp} F.~Ambrosino {\it et al.}, KLOE Collaboration, \Journal{\PLB} {638} {140} {2006} 
\bibitem{rege} A.~Di~Domenico, \Journal{\NPB}{450}{293}{1995}
\bibitem{oller} J.~A.~Oller, \Journal{\NPA}{714}{161}{2003} 
\bibitem{achasov1} N.N.~Achasov, V.~V.~Gubin, \Journal{\PRD}{64}{094016}{2001}
\bibitem{achasov2} N.N.~Achasov, V.~V.~Gubin, \Journal{\PAN}{65}{1887}{2002}
\bibitem{PDG2006} W.-M. Yao et al., Particle Data Group, {\it J. Phys.} G {\bf 33}, (2006) 1


\bibitem{feldmancousins}G.J. Feldman, R. Cousins,  \Journal{\PRD}{57}{3873}{1998}
\end{thebibliography}
\appendix
\end{document}